\documentclass{article}

\usepackage{arxiv}
\usepackage[table,xcdraw]{xcolor}
\usepackage[utf8]{inputenc} 
\usepackage[T1]{fontenc}    
\usepackage[numbers]{natbib}
\usepackage{hyperref}      
\usepackage{url}            
\usepackage{footnote}
\usepackage{booktabs}       
\usepackage{amsfonts}       
\usepackage{nicefrac}       
\usepackage{microtype}      
\usepackage{lipsum}		
\usepackage{graphicx}
\usepackage{natbib}
\usepackage{tabularx}
\usepackage{doi}
\usepackage{mathtools}
\usepackage{xcolor}
\usepackage{soul}
\usepackage{tabularx}
\usepackage{colortbl}
\usepackage{multirow}
\usepackage{multicol}
\usepackage{array}
\usepackage{amsmath}
\usepackage{booktabs}

\title{Explain to Decide: A Human-Centric Review on the Role of Explainable Artificial Intelligence in AI-assisted Decision Making}

\author{ \href{https://orcid.org/0000-0002-1464-2157}{\includegraphics[scale=0.06]{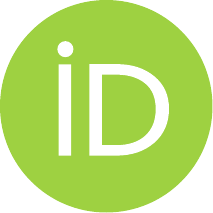}\hspace{1mm}Milad Rogha} \\
	College of Computing and Informatics\\
	University of North Carolina at Charlotte\\
	Charlotte, NC 28213 \\
	\texttt{mrogha@uncc.edu}  
}

\renewcommand{\shorttitle}{\textit{arXiv} Template}

\hypersetup{
pdftitle={Explain to Decide: A Human-Centric Review on Explainable Artificial Intelligence in AI-assisted Decision Making},
pdfsubject={A survey paper},
pdfauthor={Milad Rogha},
pdfkeywords={human-AI decision-making, explainable AI, machine learning},
}

\newcolumntype{L}{>{\centering\arraybackslash}m{0.2\textwidth}}  
\newcolumntype{R}{>{\centering\arraybackslash}X}  
\newcolumntype{C}{>{\centering\arraybackslash}m{0.08\linewidth}}
\newcolumntype{D}{>{\centering\arraybackslash}m{0.125\linewidth}X}

\newcommand{\revMilad}[1]{\textcolor{black}{#1}}

\begin{document}
\maketitle

\begin{abstract}
The unprecedented performance of machine learning models in recent years, particularly Deep Learning and transformer models, has resulted in their application in various domains such as finance, healthcare, and education. However, the models are error-prone and cannot be used autonomously, especially in decision-making scenarios where, technically or ethically, the cost of error is high. Moreover, because of the black-box nature of these models, it is frequently difficult for the end user to comprehend the models' outcomes and underlying processes to trust and use the model outcome to make a decision. Explainable Artificial Intelligence (XAI) aids end-user understanding of the model by utilizing approaches, including visualization techniques, to explain and interpret the inner workings of the model and how it arrives at a result. Although numerous research studies have been conducted recently focusing on the performance of models and the XAI approaches, less work has been done on the impact of explanations on human-AI team performance. This paper surveyed the recent empirical studies on XAI's impact on human-AI decision-making, identified the challenges, and proposed future research directions.
\end{abstract}

\keywords{decision-making \and human-AI interaction \and explainable AI}

\section{Introduction}
\label{sec:introduction}

Throughout its evolution since the 1950s, Artificial Intelligence (AI) has experienced both periods of growth and decline, known as AI springs and AI winters. However, advancements in computer hardware technology and enhanced data availability have paved the way for increased AI applications across a variety of domains, including manufacturing, healthcare, finance, management, transportation, security, education, military, and legal practice in recent years \cite{ray2019quick,duan2019artificial,abiodun2018state}.

Artificial Neural Networks (ANNs), especially Deep Neural Networks (DNNs), demonstrated outstanding performance when applied to different tasks, including optimization, pattern recognition, data trends identification, forecasting, prediction tasks and even in query processing\cite{silver2016mastering,dave2014neural,abiodun2018state, fallahian2022gan}. However, the complex, non-linear, and multilayered architecture of these models makes the internal process and the reasoning behind such outcomes challenging to understand by the end user, turning them into "black box" models \cite{kahng2017cti, Zini2022Explainability,samek2017explainable}.

Deep Neural Networks (DNNs) are an example of black-box models that are frequently used in Natural Language Processing (NLP). These models are often opaque, which means it can be challenging for users to comprehend how these models derive specific predictions or decisions. The lack of transparency in deep learning models can create a lack of confidence in their outputs \cite{danilevsky2020survey}. This absence of transparency can be particularly worrying in applications where the models' decisions carry significant consequences, such as healthcare, finance, or the criminal justice system \cite{lipton2018mythos}.

For example, the method behind the decision-making process that led AlphaGo to win a game of Go is unknown to us, but this lack of information does not have severe implications. However, in other domains, such as healthcare or autonomous vehicles, the lack of transparency and trust can be extremely costly \cite{samek2017explainable}. The absence of trust has a negative impact on their application in such domains \cite{Zini2022Explainability}. In such cases, it is crucial that end-users understand how these models work and why they produce specific results.

The use of Machine Learning (ML) models as decision-making aids for humans is becoming increasingly pervasive in domains such as medicine, public policy, and law \cite{yin2019understanding}. Although ML models outperform humans in many aspects, research shows that using ML models by laypeople may cause challenges \cite{doshi2017towards, lipton2018mythos, ribeiro2016should, caton2020fairness}. ML models, cannot be used autonomously. One of the challenges is that decision-makers may not be familiar with the uncertainty of the predictions by the ML model \cite{prabhudesai2023understanding}, leading them to over or under rely on the model outcomes \cite{antoniadi2021current}. Another issue is that the laypeople, as the end users of the ML models, may lack a clear understanding of the performance metrics of the model, leading them to misunderstand or misuse the predictions made by the model. Finally, a growing concern draws attention toward amplifying unintentional bias in predictions by the model \cite{caliskan2017semantics}.

Moreover, despite the superior performance of ML models to humans, the models cannot be applied to high-stakes tasks fully autonomously due to the probabilistic nature of the models, which means a successful outcome is not guaranteed \cite{zhang2020effect}. In these situations, the cost of a failure is catastrophic; hence, using an AI model as the sole decision-maker is not desirable \cite{lai2023towards}. Another reason that makes human presence in the loop inevitable is the ethical (and sometimes legal) concerns, e.g., in the legal system and healthcare \cite{zhang2020effect}.

As the use of DNNs has grown in recent years, so has the effort to create models that can be explained and interpreted and on which the user can rely. These efforts led to a specific branch of AI named eXplainable Artificial Intelligence (XAI). Because of the various requirements of each field, there is little consensus on the definition of XAI across all domains \cite{doshi2017towards}. Defense Advanced Research Projects Agency (DARPA) \cite{gunning2019darpa} defines XAI as "AI systems that can explain their rationale to a human user, characterize their strengths and weaknesses, and convey an understanding of how they will behave in the future ." XAI emerged as a sub-field of ML to provide human-readable explanations for three domains: the training process, the learned representations, and the model's decisions \cite{samek2021explaining}.

The growing effort to make ML models more interpretable has revealed several challenges. One of the difficulties with interpretability is the ambiguity surrounding its definition. Bibal and Frenay identified two issues defining interpretability in ML literature: first, similar terms for interpretability do not always refer to the same issue, and second, the literature is scattered and unstructured \cite{bibal2016interpretability}. They discussed in \cite{bibal2016interpretability} that interpretability can be applied to both models and their representations and that in order to measure interpretability, it is necessary to identify which specific component of the model is being interpreted. There are two approaches to measuring interpretability: the heuristic approach, which allows for quantitative measures to be applied to same-type models (for example, comparing two Support Vector Machines (SVMs) or two Random Forest models), and the user-based survey method, which is applied to model representations and allows for comparison of different types of models \cite{bibal2016interpretability}.

One goal of XAI is to increase transparency, which hopefully leads to the robustness of ML models. Even the highest-performance models may be erroneous in some cases, which makes their application with total autonomy avoidable, especially in high-stakes tasks. While models have skills such as speed, consistency, and overall accuracy, humans are able to understand nuance and context \cite{carton2020feature}. These two sets of skills, when put together, can lead to complementary team performance (CTP), i.e., the performance of a human-AI team outperforms both AI and humans\cite{bansal2021does}. The question, however, is raised: do explanations always increase human-AI performance in all cases? Does providing more information about the model enhance user decision-making performance? 

Human user study plays a vital role in understanding human behavior during the execution of a task. For instance, \cite{Doroudian2022study} designed and performed a user study to identify human behavior in search and rescue (SAR) tasks under dangerous conditions for firefighters' training programs.

The current survey paper focuses on the research around the impact of explainable artificial intelligence on AI-assisted decision-making, precisely scenarios in which humans and AI work together as a team to make a decision with stress on tasks that involve text data. The following section, section \ref{sec:application}, provide examples of the application of human-AI decision making in high-stakes situations. Section \ref{sec:xai} paints a picture of the current state-of-the-art XAI techniques, and section \ref{sec:methadology} describes the paper selection process and provides the final list of the selected papers. Section \ref{sec:decisionMaking} discuss human AI decision-making. The selected papers are reviewed in detail in section \ref{sec:xai_and_dm} with a focus on empirical studies to investigate the impact of XAI on AI-assisted decision-making performance. In section \ref{sec:discussion}, challenges for the application of XAI in AI-assisted decision-making scenarios as well as the solutions, and limitations are discussed. Finally, the paper provides future research directions in section \ref{sec:future directions}.


\section{Application}
\label{sec:application}
Humans make decisions daily with varying levels of importance. From choosing a meal from a menu to deciding whether a patient has cancer or not, a decision has to be made with the available information, often within a short time-frame. 
With the abundance of AI agents and their outstanding performance, many fields have obtained AI systems to assist decision-making. Some of these tasks are high-stakes, i.e., the cost of failure can be catastrophic, and there is no room for incorrect decisions. Although ML models outperform human decision-makers in many fields, for example, in precision and speed, humans bring unique insights that the models cannot replace in many instances. Augmenting ML outcomes with human decision-making may provide better results in these cases \cite{zytek2021sibyl}. 

 Hence, a collaboration between humans and AI happens, leading to complementary team performance, i.e., team performance above each of the agents taken individually \cite{surden2019artificial}.
This section presents some applications of human AI decision-making in high-stakes scenarios.

\subsection{Legal domain}

In recent years, the use of AI technology in legal systems has increased significantly. This technology has been leveraged to assist with a wide range of legal tasks. Acknowledging the stakeholders at play and understanding how AI is employed is essential since various actors play different roles in the legal system and use AI differently. \cite{surden2020ethics} categorized users of AI in the legal domain into three categories: Administrators, practitioners, and users. The administrators are government officials who administer law (e.g., judges, legislators, and police). Practitioners include attorneys, while users are individuals and businesses who adhere to legal regulations.

One application of AI in the legal world comes from criminal justice. It relates to the Administrators group, where judges have to make decisions about criminal defendants that appear before them to whether to release them on bail before trial or what sentence to impose if they become convicted. The common factor between both decisions is recidivism, i.e., whether the defendant may relapse to criminal activity when released.
A judge's decisions vary based on the likelihood of a defendant's recidivism. For instance, the judge may deny bail based on the high probability of a defendant committing a crime while released, or she may impose more severe sentences in a sentencing context \cite{surden2020ethics}.

\subsection{Public Health}
Another example of high-stakes scenarios of decision-making is child welfare. Child abuse is a current social issue in the United States. According to the Centers for Disease Control and Prevention (CDC), in 2018, at least one out of every seven children experienced a type of abuse, the most common being physical, emotional, sexual abuse, and neglect. In the short term, children who are victims of abuse suffer from immediate physical injuries, e.g., broken bones, cuts, or bruises. In the long term, the consequences include emotional and psychological problems, e.g., post-traumatic stress and anxiety \cite{cdc_2022}.
In 2021, child protective services (CPS) agencies received an estimated 3,987,000 referrals from concerned parties, including approximately 7,176,600 children allegedly maltreated. Of 3.9 million referrals in 2021, 51.5\% were screened and investigated, while the rest were recorded but not investigated \cite{Kelly_Street_Building2021childmaltreatment}.

Child welfare specialists evaluate these referrals to ascertain whether they merit further inquiry, an action termed as 'screen in', or whether they should be recorded but dismissed, referred to as 'screen out.' An erroneous judgment within this screening mechanism can engender grave repercussions for the implicated children, parents, nuclear and extended family members, and caregivers. In instances of a false positive, where an allegation of maltreatment is deemed credible despite its non-occurrence, the implicated children and their parents may endure persistent psychological trauma, alongside potential financial losses, career impediments, or social stigmatization. Conversely, a false negative, where actual maltreatment is erroneously dismissed, can result in the continuation of a child's suffering and, in the most severe cases, may lead to fatality \cite{zytek2021sibyl}.

While human experts deal with high volumes of referrals, there are repeated cases of missed abuse. Ignored cases of child fatality, even though referred several times yet never screened in, are rare yet exist and can be avoided. Overloaded human experts may miss the red flags, such as several referrals, which ML models can detect. An example of using ML models in a child protection context to facilitate decision-making is the application of predictive risk modeling (PRM). PRMs have been deployed in several counties, including the model deployed in Allegheny County, PA \cite{vaithianathan2017developing}.

\subsection{Content Moderation}
The growth of social network platforms has led to a change in how content is moderated, shifting it from traditional community moderation \cite{lampe2004slash} toward what is known as "platform moderation" or "commercial content moderation" \cite{roberts2018digital}. A large body of research identifies challenges that commercial content moderation faces, including labor concerns (e.g., the mental burden on the moderators), how democratic the moderation is since a specific group is in charge of setting the rules, and overall opacity and accountability of the process \cite{gillespie2018custodians, kaye2019speech, suzor2019we, roberts2018digital}.

Instances such as the live broadcast of the Christchurch mass shooting in March 2019 serve as notable examples that highlight the significance of content moderation. During the Christchurch mass shooting incident, the perpetrator utilized the Facebook platform to livestream the entirety of the attack. Despite the expeditious removal of the initial video, approximately 1.5 million replicas were generated and subsequently disseminated across multiple platforms. Nevertheless, about 80\% of these duplicates were automatically intercepted and barred, thwarting their upload. The Christchurch incident posed a significant challenge for the Global Internet Forum to Counter Terrorism (GIFCT) \cite{gorwa2020algorithmic}. GIFCT was established in 2017 by prominent technology companies such as Facebook, Microsoft, Twitter, and YouTube, with the aim of curbing the dissemination of terrorist and extremist material across online platforms \cite{GIFCT_2019}.

Major technology companies that own social media platforms are facing mounting pressure from official regulations to enhance their content moderation practices. Platforms are required to expeditiously remove illegal or problematic content within a limited timeframe, as stipulated by regulations such as the German NetzDG or the EU Code of Conduct. Consequently, there has been an increasing endeavor to explore technological approaches in the field of content moderation, aiming to offer efficient and large-scale solutions \cite{gorwa2020algorithmic}.

One strategy to automate content moderation is to use AI models. The use of machine learning models for toxic comments and hate speech detection has raised the research community's attention in recent years.
Ayo et al. \cite{ayo2020machine}  categorized the ML models used for content moderation into single and hybrid models. Examples of the single method include the fuzzy multi-task learning (FML) developed by Liu et al. \cite{liu2019fuzzy} to detect cyber hate. The developed method is a trifold. The first step classifies the tweet into hate/non-hate. The second step identifies the type of hate (religion, race, sexual orientation, disability), and the final step detects the topic and the context of the tweet. Other examples include the use of Bayesian Networks \cite{chakravartula2019hateminer, khond2020preventive, kiilu2018using}, SVMs \cite{vega2019mineriaunam, perello2019ua}, and Long Short-Term Memory (LSTM) architecture \cite{nguyen2019vais, paschalides2020mandola} . 

The hybrid method combines different ML models to achieve an enhanced performance compared to the baseline performance of each model alone \cite{ayo2020machine}. Examples of hybrid methods include Bayesian network and recurrent neural network (BN-RNN) \cite{rother2019german}, LSTM and neural networks (A combination of LSTM and other deep learning architectures) \cite{wang2019ynu, liu2018neural}, and Naive Bayes and Logistic Regression \cite{miok2019prediction}. Fore more variations of state-of-the-art ML models applied to content moderation on Twitter data see \cite{ayo2020machine}.

\section{Explainable Artificial Intelligence}
\label{sec:xai}
The performance of ML models has been rapidly increasing over recent years. The superior performance stems from the complexity of the model architectures, which comprises millions of parameters, e.g., in DNNs \cite{arrieta2020explainable}. Due to the high performance of the recent models, their application in various real-world problems has become ubiquitous, including healthcare \cite{esteva2019guide, miotto2018deep}, finance \cite{heaton2017deep, huang2020deep}, and legal \cite{huang2021context, wei2018empirical} domains. However, this increased performance has come at the cost of model complexity, which results in reduced model transparency. An example of this is the Deep Learning (DL) paradigm, which is the heart of many today's models. The lack of transparency in these complex model architectures consequently turns them into black-box models. Black box models include DNNs and ensamble or hybrid models (see section 2 for examples of hybrid models) \cite{linardatos2020explainable}.

\begin{figure}[!htb]
    \centering
    \includegraphics[width=0.6\linewidth]{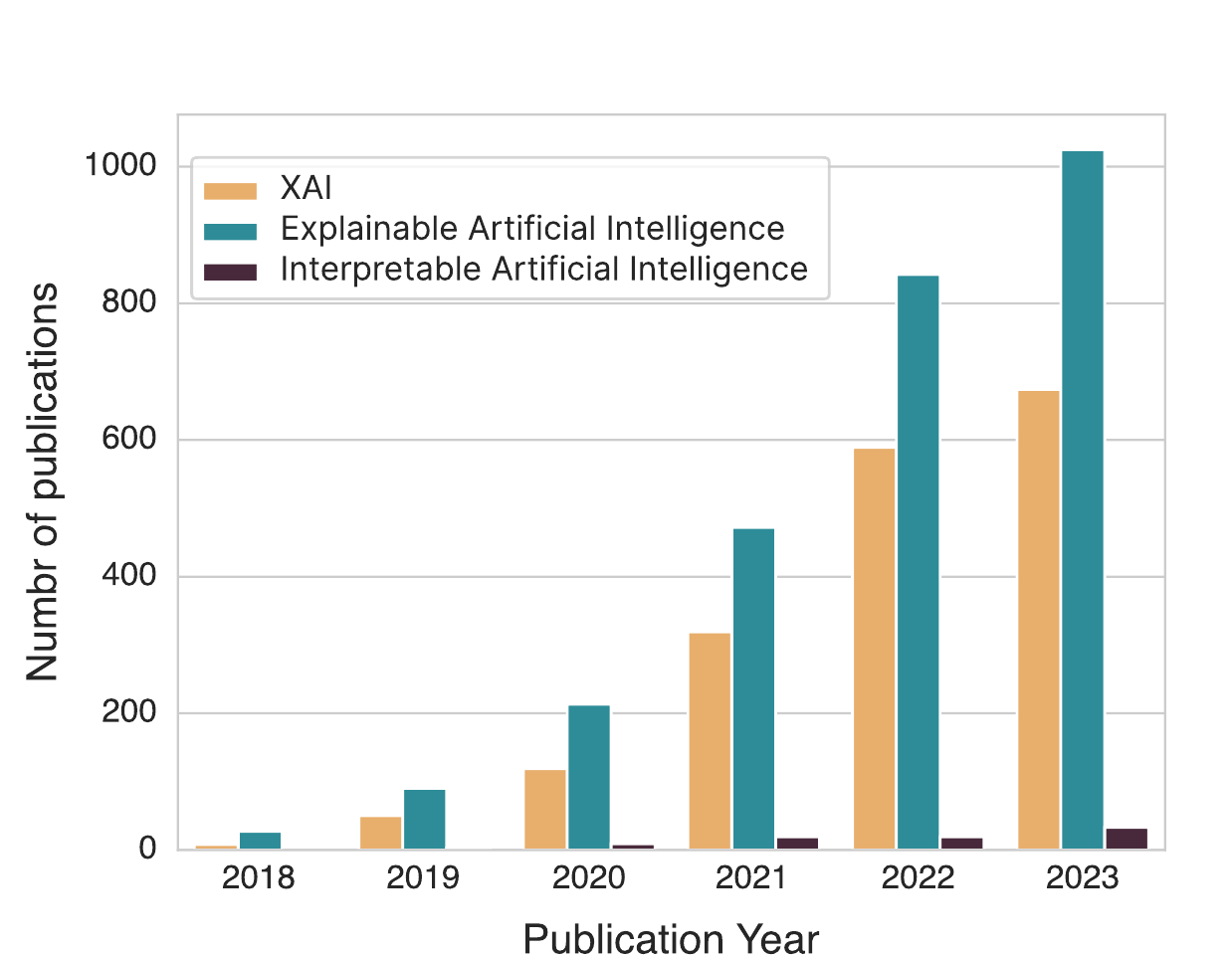}
    \caption[XAI trend]{The number of papers published between 2018 and 2023 whose title or keywords contain any of the phrases XAI, Explainable Artificial Intelligence, or Interpretable Artificial Intelligence.
    The data is retrieved using Scopus developer's API. \footnotemark }
    \label{fig:xai_trend}
\end{figure}

Humans are hesitant to adopt technologies that lack direct interpretability \cite{zhu2018explainable}. The lack of interpretability and comprehension about how models work contributes to a lack of trust in those models. When humans do not trust a model, they may not use it. 
The use of Machine Learning (ML) models in situations where they make important decisions and predictions has led to increasing demand from stakeholders for greater transparency of the models. Out-performing Artificial Intelligence (AI) systems, especially ML models applied to real-world scenarios, require transparency, trustworthiness, fairness, and robustness. This need has led to the development of eXplainable Artificial Intelligence (XAI), a sub-branch of ML that focuses on understanding and interpreting the mechanism and behavior of AI systems \cite{gunning2019darpa}. Although the research community was primarily focused on evaluating the predictive power of AI models, the field of explainable artificial intelligence (XAI) had been overlooked \cite{linardatos2020explainable}. However, in recent years, there has been a growing interest in XAI, as evidenced by the increasing number of academic contributions in this field illustrated in Figure \ref{fig:xai_trend}.

\footnotetext{\url{https://dev.elsevier.com/documentation/ScopusSearchAPI.wadl}}

\subsection{Interpretation vs. Explainability and Explanation}

The terms "explainability" and "interpretability" have often been used interchangeably by researchers. Nevertheless, considerable efforts have been undertaken to differentiate these two concepts and establish metrics for assessing both of them. The issue of establishing universally accepted terminology remains a persistent challenge \cite{adadi2018peeking}. Lipton identifies that the problem with the definition of interpretability stems from a sparse methodology and objective in investigating interpretability \cite{lipton2018mythos}. Doshi-Velez and Kim defined interpretability as "the ability to explain or to present in understandable terms to a human" \cite{doshi2017towards} while finding a formal definition for explainability evasive. Biran and Cotton \cite{biran2017explanation} considered a system interpretable if "their operations can be understood by a human." Miller adapted their definition and defines interpretability as a criterion of how well a human could understand the model decision, i.e., model outputs, in a given context and explanation as "explicitly explaining decisions to people" \cite{miller2019explanation}.
Gilpin et al. \cite{gilpin2018explaining} defined explanation as the answer to "why" questions and noted that explanations can be evaluated based on their interpretability and completeness. Interpretability aims to describe the workings of a system in a way that is understandable to humans, while completeness accurately describes the mechanics of a system. For instance, a complete explanation of a Deep Neural Network (DNN) should reveal all internal parameters and the mathematics behind the system. The authors argued that interpretability alone is not enough for humans to trust black box models and that explainability is necessary to summarize the reasons for neural network behavior, gain user trust, and explain why the model made certain decisions.

Hence, interpretability can be inferred as answering "why?" about the output, i.e., understanding the relationship between its inputs and outputs. In contrast, explainability answers "how?", i.e., the understanding of the internal procedures behind a model's output \cite{linardatos2020explainable}.

\subsection{Necessity of XAI}
In situations where the consequences of unacceptable results are inconsequential, or the problem itself is validated in real-world applications in a way that we trust the model regardless of its defects, the explanations may not be necessary. Examples include the application of ML systems in music recommendation, postal code sorting, and aviation collision systems. On the other hand, in their paper \cite{doshi2017towards}, Doshi-Velez and Kim argue the necessity for interpretability due to what they call problem formalization incompleteness. Incompleteness can introduce unquantifiable bias to the model. Examples of problem incompleteness, as stated in \cite{doshi2017towards}, include scientific understanding, safety, ethics, and mismatched objectives.
Human beings seek knowledge, and the best way to gain knowledge is to transform explanations into knowledge. Regarding safety, many real-world situations are too complex to be thoroughly tested, e.g., autonomous driving. Considering all possible outputs to identify where the model could err is impossible. That is where explanations become beneficial to provide an understanding of the model decisions.
Ethically, the systems should be protected against unwanted bias and discrimination. In many situations, having a fair model is crucial, such as loan approval, fraud detection, and hiring. A model is prone to various types of bias—for example, unwanted gender-based bias in creating word embedding \cite{doshi2017towards} or recidivism \cite{hamilton2019sexist}.

\subsection{Evaluation of Interpretability}
The evaluation metrics for gauging a system's performance vary based on the nature of the work. When evaluating a system for a specific task, we should consider how successful it is in performing that task, e.g., an ML model beat a human player in chess. On the other hand, when evaluating a system as a whole, we need to measure its generalizability by testing it on various benchmarks. Doshi-Velez and Kim \cite{doshi2017towards} laid out a threefold evaluation taxonomy for interpretability: application-grounded, human-grounded, and functional-grounded.

Application-grounded evaluation measures how successful a system is in a specific application. In other words, this type of evaluation compares the system's performance to a task-specific benchmark within the specific application. This evaluation usually includes conducting human experiments. The quality of explanations is evaluated based on their effectiveness in achieving the end task, such as reducing discrimination or enhancing error identification. Additionally, the benchmark for this evaluation is the ability of human-produced explanations to help other human subjects in completing a task.

In the human-grounded evaluation, a simpler version of the application is used. This makes the human-subject experiment more straightforward and easier to conduct without the need for experts. As a result, the experiment can be conducted with lay users, potentially leading to a larger pool of subjects. The primary focus of the human-grounded evaluation is not the end task but rather to ensure that the explanations produced are successful in conveying the general notions to the human subject. Examples of potential experiments for human-grounded evaluation are Binary Forced choice, Forward simulation/prediction, and Counterfactual simulation. Binary forced choice is discussed in section \ref{sec:decisionMaking}.

Functionally grounded evaluation replaces human experiments with a formal definition of interpretability as a proxy to evaluate the quality of explanations. This type of evaluation is most beneficial when going through necessary approvals (e.g., IRBs) are not feasible for the researcher, the method for evaluation is still in development, or there exist ethical concerns for human subject experiments.

Monlar gathered a set of properties for explanation methods and the explanations themselves that could result in good explanations. Properties for explanation methods consist of Expressive Power, Translucency, Portability, and Algorithmic complexity. Properties for a particular explanation include Accuracy, Fidelity, Consistency, Stability, Comprehensibility, Certainty, Degree of Importance, Novelty, and Representativeness \cite{molnar2020interpretable}.

According to recent research \cite{jacovi-goldberg-2020-towards, ding-koehn-2021-evaluating}, the faithfulness of a method is related to the consistency of the rationale and the input/output provided. Wang et al. used a similar mechanism to assess model faithfulness by assessing the consistency of the explanations provided in response to perturbations \cite{wang2022fine}. The selection of appropriate measures and metrics is a case-dependent issue, and finding a solution remains an ongoing challenge \cite{linardatos2020explainable}.

\subsection{Taxonomy of XAI}
A large body of work provided taxonomies for XAI evaluation, methods, and techniques \cite{xu2019explainable, adadi2018peeking, gilpin2018explaining, mohseni2021, linardatos2020explainable, arrieta2020explainable, alicioglu2022}. This paper does not aim to provide an extensive list of taxonomies for XAI.  Instead, first, a common taxonomy of techniques based on the literature is provided. The rest of the section then focuses on XAI techniques that have been applied to decision-making tasks using text data.

Gilpin et al. \cite{gilpin2018explaining} examine explainable models for deep learning in their work. They propose a taxonomy that categorizes the explanations provided by these models into three categories. The first category explains the relationship between the input and output of the network, answering why a particular output is achieved from a given input. The second category attempts to explain how data is represented inside the deep network. The last category includes transparent networks, which are networks that can explain themselves. The authors reviewed the related work and suggested evaluation metrics for each category.

Adadi et al. conducted a thorough review of the field of XAI and proposed a taxonomy based on four main axes. They emphasized the need for more formalism in the field and highlighted the current trends and methods being used. Their review also brought to light the fact that there is a lack of research on the impact of human factors in XAI. Additionally, they provided suggestions for future directions and recommended the combination of existing methods for future research \cite{adadi2018peeking}.

Linardatos et al. \cite{linardatos2020explainable} have developed a taxonomy of interpretability methods for machine learning (ML) based on their purpose of interpretability. They have classified the methods into four categories. The first category includes methods that interpret pre-trained black box models. The second category includes methods that aim to generate white box models that are interpretable and easy to understand for humans. The third category includes methods that enhance fairness. Finally, the fourth category includes methods for models' sensitivity analysis. These methods involve testing the sensitivity of predictions to intentional input changes. Doing so helps to test the stability and trustworthiness of the predictions.

Another typical dimension to separating interpretability methods is whether they provide local explanations, i.e., explaining a single prediction (outcome), or global, meaning that they explain the whole model. 

Furthermore, methods can only be applied to specific types of algorithms or models, making them model-specific. In contrast, methods may be applicable to any models and algorithms, making them agnostic to the model itself. These methods are called model agnostic.

\begin{figure}[htb]
    \includegraphics[width=\textwidth]{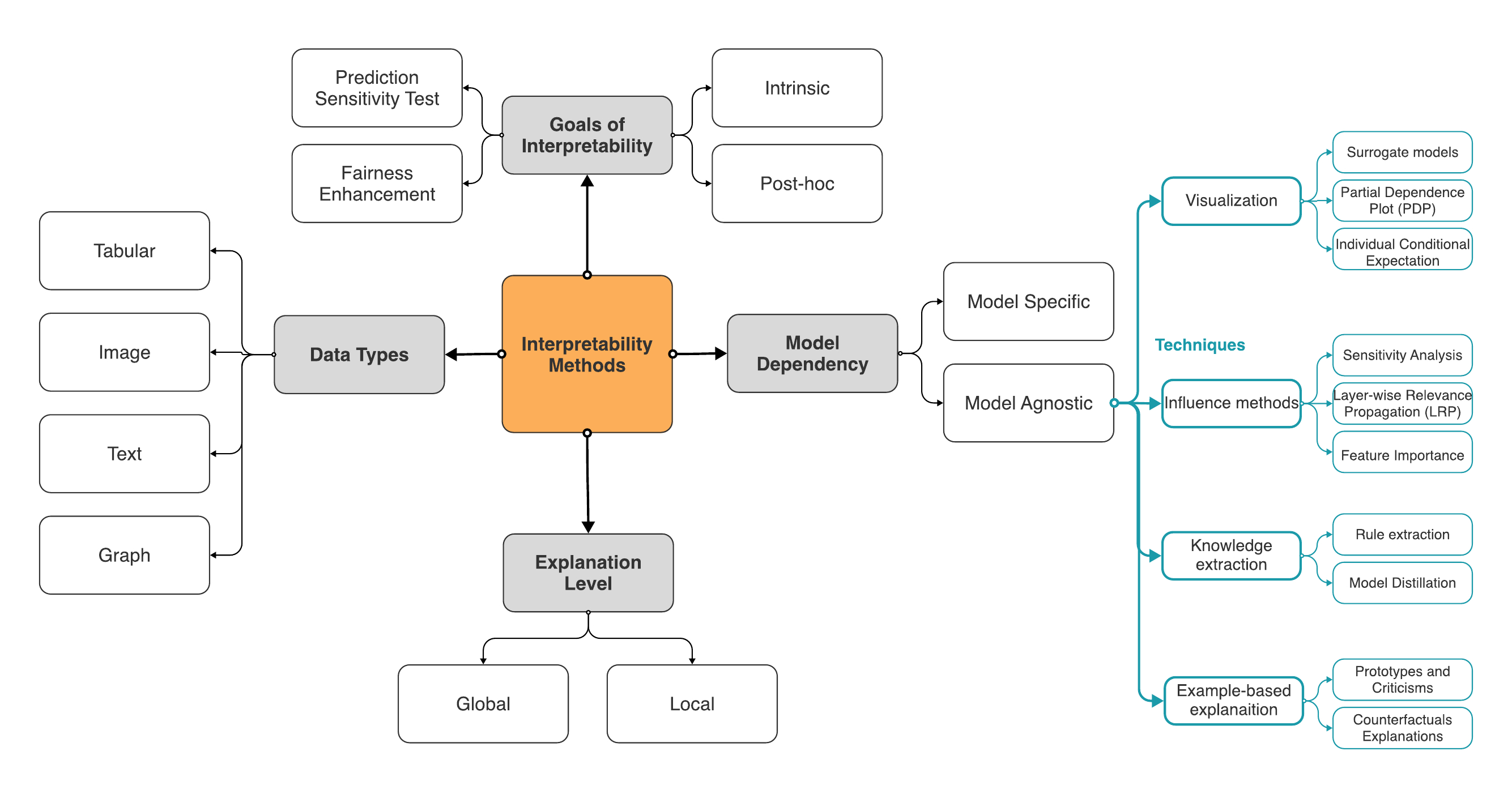}
    \caption{A taxonomy XAI methods inspired by the work of \cite{linardatos2020explainable, adadi2018peeking}}
    \label{fig:xai_taxonomy}
\end{figure}

Finally, while some methods are only applicable to a specific type of data, e.g., image, tabular, text, or graph, others work with two or more types of data. Given all these dimensions, figure \ref{fig:xai_taxonomy}, inspired by the work of \cite{linardatos2020explainable, adadi2018peeking}, illustrates a taxonomy of the interpretable methods.

\section{Methodology}
\label{sec:methadology}
The selection of papers for this survey began with looking for topics including human-AI collaboration in decision-making using the explainable artificial intelligence method. Specifically, the survey looked for papers that included human experiments conducted to find the impact of XAI on human decision-making.  

First, a keyword-based search using keywords including combinations of "AI-assisted," "Human-AI," "decision making," "XAI" decision making," and "Human-computer decision making" and their equivalents were conducted. Academic databases used include Google Scholar, ACM Digital Library, and IEEE Xplore search platforms to search for and collect papers. Due to the young age of the XAI field and its rapid growth and development, preprints from arXiv were also included.  

Moreover, to include any related contribution, the references section of each paper was searched for relevant papers that may have been missed during the search process.
To narrow down the search, only papers published from 2019 and onward are included, with the exception of foundational works that are necessary to define a concept or provide context.  

Using the Google Scholar search platform, the search query \textit{"AI-assisted decision making" AND "explainable artificial intelligence" AND "empirical study"} retrieved 135 papers. Adding the publication year filter reduced the results to 123 papers, 17 of them from ACM. \revMilad{A refined version of the query, i.e., \textit{( (AI-assisted" OR "human-AI" ) AND "decision making" ) AND ( "explainable artificial intelligence" OR "XAI" OR "Interpretable AI" ) AND ( "empirical study" OR "user study" OR "case study" )}) resulted in 21 papers}. Query on the Scopus database resulted in 7 papers, raising the total number of papers to 130. The query used is: \textit{'TITLE(AI-Assisted AND Decision Making) AND TITLE(XAI OR Explainable Artificial Intelligence) OR KEY(AI-Assisted Decision Making) OR ABS(decision making AND XAI OR explainable artificial intelligence)'}. Due to the high number of papers, an exhaustive review was out of reach. Hence, eight papers that include empirical studies on the impact of XAI on human-AI decision-making were selected. Other factors that contributed to the paper selection include the higher impact and recentness of the selected papers.

Table \ref{tab:decisionMakingEmpiricalStudies} provides a comprehensive list of the final selected papers.

\begin{table}[!ht]
    \small
    \caption{List of selected papers surveyed.}
     \label{tab:decisionMakingEmpiricalStudies}
    \begin{tabularx}{\linewidth}{X C C }
      
    \toprule 
        \textbf{Title} & \textbf{Reference} & \textbf{Publication Year} \\
        \midrule
         \addlinespace
        Effects of Explainable Artificial Intelligence on trust and human behavior in a high-risk decision task & \cite{leichtmann2023effects} & 2023 \\
         \addlinespace
        Measuring appropriate reliance in human-AI decision-making & \cite{Schemmer2023AReliance} & 2023\\
         \addlinespace
        Explanations can reduce Overreliance on AI systems during decision-making & \cite{vasconcelos2023explanations} & 2023 \\
        \addlinespace
        Human-AI collaboration via conditional delegation: A case study of content moderation & \cite{lai2022human} & 2022 \\
        \addlinespace
        Does the whole exceed its parts? the effect of AI explanations on complementary team performance & \cite{bansal2021does} & 2021 \\
        \addlinespace
        Does explainable artificial intelligence improve human decision-making? & \cite{alufaisan2021does} & 2021 \\ 
        \addlinespace
        Feature-based explanations don't help people detect misclassifications of online toxicity & \cite{carton2020feature} & 2020\\
         \addlinespace
        Effect of confidence and explanation on accuracy and trust calibration in AI-assisted decision making & \cite{zhang2020effect} & 2020 \\
        \addlinespace
        \bottomrule
    \end{tabularx}

\end{table}

      


\section{Decision Making}
\label{sec:decisionMaking}
Humans are faced with making decisions on a daily basis, some of which can have significant consequences in areas such as criminal justice, finance, and healthcare. While Machine Learning (ML) models, particularly Neural Networks (NNs), have shown promising results in many fields and situations, there is hesitation towards fully automating decision-making in these areas due to ethical, safety, and legal concerns \cite{lai2022human}. However, manual handling of these situations, whether it be deciding on a patient's cancer\cite{sarwinda2021deep}, approving a loan \cite{binns2018s}, or determining probation, is time-consuming and susceptible to human error.
Extensive research has been conducted in the field of human-AI interaction due to the problems mentioned above. This section provides an overview of human-AI decision-making, followed by a survey of recent empirical studies on human-AI decision-making from a selection of papers. The studies are categorized based on the decision-making task, study design, and metrics used, and the positive and negative aspects of each category are discussed.

\subsection{Related Work}
In their survey \cite{lai2023towards}, Lai et al. surveyed papers that include empirical studies on human decision-making and identified the current trends and gaps, as well as provided future suggestions that help reach a design framework for human-AI decision-making studies. They surveyed papers that focused on the human decision-maker rather than the AI component or other AI stakeholders, such as ML experts. In addition, the papers should include human subject experiment to be considered, resulting in the exclusion of qualitative studies. Finally, for a paper to be included in their scope of review, Lai et al. focused only on the studies that had a classification or regression task, excluding papers with sequential decision-making, games, and other sorts of decision-making tasks \cite{lai2023towards}. They reviewed the papers based on the task, the metrics for evaluation, and the type of AI assistance. The task itself was then broken down based on the domain, level of risk (low, high, and artificial tasks with little to no risk), and subjectivity of the task. The AI assistance type includes providing information about the prediction or the model and uncertainty. The evaluation metrics were categorized into two categories with respect to the decision-making task and the AI. The former group includes metrics such as efficacy, efficiency, task-level satisfaction, and mental demand.
In contrast, the latter includes an understanding of the model, trust, fairness, and system satisfaction and usability. They identified that the lack of a coherent study design framework, both in terms of the task and in terms of the evaluation metric, results in a dispersed list of studies with results that may not be necessarily generalizable. One reason behind this is that the tasks in the studies they surveyed vary in domains. Hence, each had different risk levels, required different metrics, and varied in the amount of expertise required from the participant \cite{lai2023towards}.

Schemmer et al. \cite{schemmer2022meta} conducted a meta-analysis, reviewing 44 studies from 9 papers that met their inclusion criteria. Their findings suggest that the presence of XAI algorithms in conjunction with AI components, as opposed to sole AI output without any additional information, did not show a significant difference in performance. However, when XAI assistance was present, there was an overall improvement in human task performance. Additionally, their analysis revealed that XAI is more effective when applied to text data compared to tabular data. The authors highlight the ambiguity of previous studies and encourage future research on various data types and different tasks based on the task complexity and the level of human knowledge of the task.

Next, a brief overview of decision models proposed by psychologists is provided. Psychologists proposed decision-making models to represent the human decision-making process mathematically. Understanding these models enhances the design of human-AI decision-making systems by providing a better perspective on how humans decide and in the identification of human factors involved in the process to better account for them in designing human-in-the-loop systems. While a comprehensive review of these models is out of the scope of this paper, a selection of decision-making models are discussed in this section.

\subsection{Decision Making Models}
Everyday, individuals are faced with the task of making forced-choice decisions, wherein they are required to select from a set of alternatives, taking into consideration the value of a specific variable of interest. Examples of this type of decisions range from decisions with minimal consequences, such as choosing a film to view for the evening, to decisions of significant importance, such as determining the presence or absence of malignancy in a patient \cite{lee2004evidence}.

A popular method for modeling how individuals make decisions is to maximize the usefulness of the decision by taking into account all relevant information available without considering the efficiency of the decision-making process. This approach is commonly known as the "rational approach" or "substantively rational," as referred to by Simon \cite{simon1976substantive}. It is a key theory in decision-making and is widely used in both physical and behavioral sciences. Over time, many successful models have been made based on the rational model \cite{lee2004evidence}.

Another approach to modeling decision-making is the \textit{Fast-And-Frugal} approach. In developing this approach, Gigerenzer and Todd \cite{gigerenzer1999fast} argued that due to the nature of decision-making in competitive situations where time is of the essence, the decision-making process must be time-efficient. This approach accents the role of the environment in shaping human decision-making and the interaction between mental processes and the external environment of the task. Based on their argument, given that the first piece of information in a stimulus can predict other pieces of information and exploring other pieces of information requires significant effort, we can only consider the first piece of information. The same logic also holds for a situation in which each further piece of information provides less information than the previous pieces. Gigerenzer and Todd developed several models based on the Fast-And-Frugal approach. One of these models is the Take The Best (TTB) model of forced choice.

\subsubsection{Take The Best Model of Decision Making (TTB)}
The TTB model developed by Gigerenzer and Todd \cite{gigerenzer1999fast} represents stimuli by the presence or absence of a set of properties called cues and their validities. A cue's validity is the rate at which it makes correct decisions when the cue distinguishes between the two alternatives. Equation \ref{equation:vue_validity} formally describes the validity ${v_i}$ for the ${i}$th cue:

\begin{equation} \label{equation:vue_validity}
    \centering
    v_i \equiv p(A>B  |  a_i=1, b_i=0)
    \end{equation}

While Gigerenzer and Todd \cite{gigerenzer1999fast} adopted a frequentist approach to describe cue validity, Lee et al. \cite{lee2019using} used a Bayesian approach to overcome the shortcomings of the frequentist approach to cue validity. The frequentist approach is not sensitive to the rate at which a cue discriminates between the alternatives. For example, there is no difference between a cue that gets its only decision correct and a cue that gets all its 150 decisions correct; both get a validity of 1. On the other hand, in the Bayesian approach, the cue with only one decision made gets a validity of ${\simeq .67}$, while the other cue with 150 out of 150 correct decisions gets a validity of ${\simeq .99}$. Table \ref{tab:ttb_rat_cueVal} exhibits the equations for each approach.

The TTB decision model evaluates the validity of each cue and examines the one with the highest validation for a pair of stimuli. A decision is made based on how the cue discriminates between the stimuli. If the cue cannot discriminate, the TTB model examines the next cue with the highest validation. The process continues until a decision is made or all cues are exhausted; in this case, a random guess is made \cite{lee2004evidence}.

\begin{figure}[!htb]
    \centering
    \includegraphics[width={.8\textwidth}]{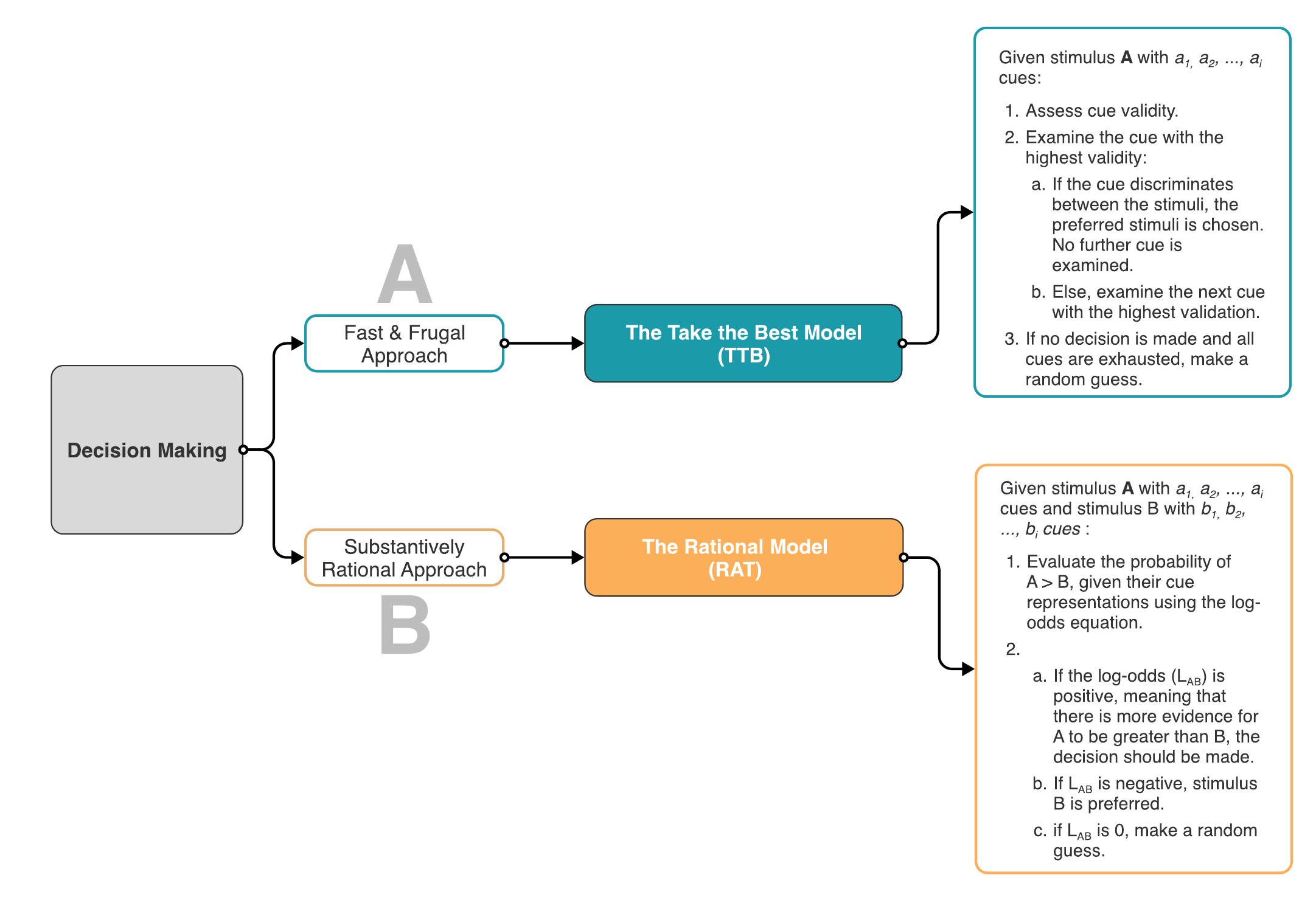}
    \caption{A high level illustration of Take The Best (TTB) and Rational model (RAT) of decisions. For equations of each model see Table \ref{tab:ttb_rat_cueVal}}
    \label{fig:decisionMaking_models}
\end{figure}

\subsubsection{The Rational Model of Decision Making (RAT)}
The Rational Model (RAT) is based on the concept of using all relevant available information to make a decision. In order to decide Stimulus A over Stimulus B, one way is to calculate and decide by the probability that Stimulus A is greater than Stimulus B. If the calculated probability is greater than 0.5, then Stimulus A is chosen over Stimulus B. In contrast, if the probability is less than 0.5, the other stimulus, Stimulus B, is chosen. If the probability is exactly 0.5, a guess has to be made. \cite{lee2004evidence} provides a simplified equation to calculate the probability that Stimulus A is greater than Stimulus B by their cue representation using Bayes' theorem (see Table \ref{tab:ttb_rat_cueVal}B). The equation is based on cue validities. Positive values of $ L_{AB} $ indicate that more evidence by the cues is in favor of Stimulus A, while negative values indicate the opposite. A random choice is made when $ L_{AB} = 0 $ \cite{lee2004evidence}.

\begin{table}[htb]
\small
\centering
\caption{Take-The-Best (TTB) and Rational (RAT) decision models }
\label{tab:ttb_rat_cueVal}
\setlength{\arrayrulewidth}{0.2mm}
\renewcommand{\arraystretch}{1.8}
\begin{tabularx}{.7\textwidth}{LR}
\hline
\multicolumn{2}{c}{\cellcolor[HTML]{dddddd}\textbf{A. The Take The Best Model (TTB)}} \\ \hline
\textit{Approach}                 & \textit{Cue validity}\\
\cmidrule{1-2}
\small
\multirow{2}{*}{Frequentist}      &  \multirow{2}{*}{$\begin{aligned}\hat{ v_i} = \displaystyle\frac{ n_{\text{correct decisions made by } i\text{th cue}}}{n_{\text{decisions made by } i\text{th cue}}}\end{aligned}$} \\
\small\\
\multirow{2}{*}{Bayesian}         & \multirow{2}{*}{ $\begin{aligned}\hat{ v_i} = \displaystyle\frac{ n_{\text{correct decisions made by } i\text{th cue}} + 1}{n_{\text{decisions made by } i\text{th cue}} + 2}\end{aligned} $ }\\
& \\
\hline
\multicolumn{2}{c}{\cellcolor[HTML]{dddddd}\textbf{B. The Rational Model (RAT)}}      \\
\hline
\small
\multirow{2}{*}{Log-Odds}         & \multirow{2}{*}{$ L_{\text{AB}} = \displaystyle\sum_{i\ni \text{FA}} \ln{\frac{\hat{v}_i}{1- \hat{v}_i}} - \displaystyle\sum_{i\ni \text{FB}} \ln{\frac{\hat{v}_i}{1 - \hat{v}_i}} $} \\     & \\           
\end{tabularx}%

\end{table}

\subsubsection{Unified model based on Evidence Accumulation}
\begin{figure}[!htb]
    \centering
    \includegraphics[width={\textwidth/2}]{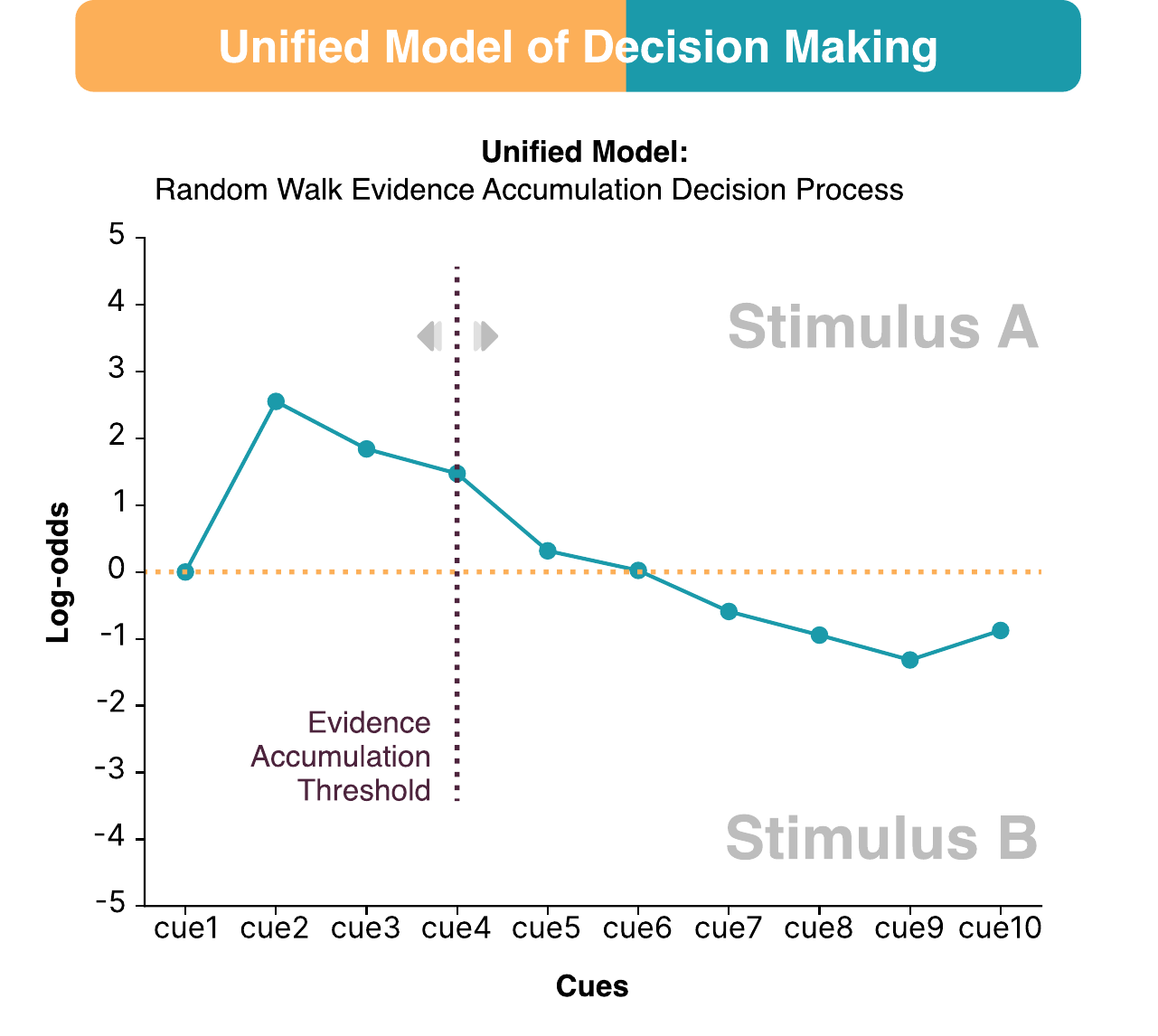}
    \caption{In a unified model, random walk evidence accumulation is applied to examine cues from highest to lowest validity. At each cue, the state of the random walk is updated based on the evidence provided. When a stimulus has a cue, it updates the random walk to move toward that stimulus. The random walk process is terminated by the threshold set for the decision. For instance, in this example, a large threshold value would cause Stimulus B to be chosen. For more information, please refer to \cite{lee2004evidence}.}
    \label{fig:decisionMaking_randomWalk}
\end{figure}

Lee and Cummins proposed a model of decision-making that unifies TTB and RAT models based on evidence accumulation \cite{lee2004evidence}. The idea behind their proposed model is based on treating both TTB and RAT models as sequential sampling processes in which the TTB arrives at a decision as soon as it finds the first evidence in favor of one of the choices, and the RAT makes a decision when all the available evidence is examined. Sequential sampling models provide an understanding of both the speed and the accuracy of performance. These models are based on the concept that the representation of stimuli in the central nervous system is changing (noisy). To make a decision, a certain quantity of evidence is needed, which is obtained by accumulating a successive number of samples of evidence (noisy representations of the stimulus) \cite{ratcliff2004comparison}. Using a specific type of sequential sampling known as a \textit{random walk}, the proposed model by Lee et al. unifies the contrasting TTB and RAT models as special cases of the random walk. Figure \ref{fig:decisionMaking_randomWalk} shows the state of the random walk after assessing the evidence of each cue. Cues are assessed in order from the highest validity to the lowest. Based on a set threshold for the number of cues, the random walk considers each cue. Based on which stimulus the evidence provided by each cue favors, the random walk moves toward that stimulus. Based on the chosen threshold, the TTB model corresponds to a low evidence threshold, while RAT corresponds to a high evidence threshold.

\subsubsection{Random Walk vs. Accumulator Model}
Due to the sequential structure of documents, text classification is a non-compensatory decision-making process that does not require reading all the words. Lee and Corlett \cite{lee2003sequential} developed two models to capture document text classification: one based on the random walk and the other based on accumulator sequential sampling processes. While both methods establish a threshold to make a decision, The difference between the random walk and accumulator sequential sampling is that in Random Walk, where there are two choices, each piece of evidence updates the accrued evidence total toward a stimulus. When the threshold is reached, a decision can be made based on the final state of the evidence total. However, the accumulator model maintains a separate evidence total for each stimulus. In their experiment, the participants were asked to identify whether a document was about the given topic through a serialized display of words from the document.
The authors defined two sets of instructions. A group of participants were asked to make decisions as quickly as possible, while the other group was asked to make decisions as accurately as possible. The results showed that these two conditions of speed and accuracy did not affect the distribution of choices, similar to confidence in the decisions made. However, under the accuracy condition, participants read twice as the words read in the speed condition, suggesting that there is not necessarily a trade-off between speed and accuracy in decision-making. The authors also fitted the random walk and accumulator models to the empirical data and compared the implemented models to several ML models. Finally, one of the shortcomings of the random walk model is that every piece of evidence may dissolve the effect of the previous evidence. For example, when the new evidence is in favor of Stimulus A, but the previous stimulus was in favor of Stimulus B, the new evidence decreases the total and hence makes reaching a threshold harder \cite{lee2003sequential}.

\subsection{Reliance}
During the process of making a decision, humans tend to seek advice. Medical and financial situations are real-world examples in which individuals seek advice. In recent days, with the prominent presence of AI and its superior performance in comparison to humans in most cases, machine-based advisors are becoming prevalent. Although research revealed that AI advice is not always beneficial, the predominating utilization of AI-based advice with high-performance models leads to a over-reliance by the human end-user, especially in cases where the human subject is not able to distinguish the correctness of the advice and hence fails to ignore the incorrect advice. Considering that the decision-maker can develop the AI-based advisor due to their requirements, following all AI advice may lead to unwanted outcomes \cite{Schemmer2023AReliance}.

\subsubsection{Appropriate Reliance}
The definition of Appropriate Reliance (AR) varies across related literature \cite{wang2008selecting, Schemmer2023AReliance}. Although in the fundamental work of Lee and See \cite{lee2004trust} on trust in automation, they did not provide a definition of AR, and their focus was on trust from various perspectives and the factors that affect its appropriateness, they provided examples of inappropriate AR, which can be costly and catastrophic.
Trust guides reliance\cite{lee2004trust} on automation due to people's social response to technology and due to the possible similarity between human reactions to computers and human reactions to human collaborators \cite{reeves1996media}. Lee et al. provided examples of inappropriate reliance, such as \textit{misuse} and \textit{disuse}. Misuse refers to situations in which accidental violation of critical assumptions led to inappropriate reliance and hence resulted in failures. In contrast, disuse refers to failures as the result of under-reliance, both examples of inappropriate reliance that can cause negative consequences \cite{lee2004trust}.   
Other research defines Inappropriate reliance as \textit{under} or \textit{over} reliance \cite{lu2021human}. The literature describes AR in a case-by-case manner. Wang et al. \cite{wang2008selecting} warned against drawing conclusions when analyzing reliance measures using indicators that lack a benchmark and emphasized the importance of defining optimal reliance. It is essential to consider the individual relative accuracy of humans and automation to decide on the appropriateness of a reliance. For example, in an extreme case, total reliance may even be beneficial if (in this case) automation outperforms humans significantly. The authors categorized the reliance indicators used in the previous literature into four paramount groups: consistency, performance, behavior, and response bias indicators. On another point, the authors explained how the specificity of reliance is essential in defining optimal reliance. Specificity of reliance refers to whether the evaluation concerns the reliance of the participant (human) on the whole or a specific type of automation feedback. For example, when studying the reliance on a system that distinguishes the friendly from the enemy, it is crucial to discriminate between reliance on "friendly" feedback versus "unknown" feedback, as the latter does not guarantee a friendly soldier. All the mentioned indicator groups above are capable of evaluating reliance on the whole system, while all except consistency indicators are suitable for indicating specific reliance \cite{wang2008selecting}.

Previous research focused on maximizing the amount humans follow the advice (reliance). However, the focus has changed recently due to the increased applications of imperfect AI advisors, more alignment between AI advisors and human decision-makers, and the potential for complimentary team performance \cite{Schemmer2023AReliance}.

\subsubsection{Reliance Metrics}
The related literature suggests a lack of a universal definition of appropriate reliance and metrics to measure AR quantitatively. Appropriate reliance on AI advice is often measured as the percentage of following the correct AI advice and the percentage of ignoring the incorrect AI advice \cite{bansal2021does}. This measurement approach, however, has a drawback in clarifying whether the human decision-maker correctly discriminated and chose the correct AI advice or whether it was simply an overlap of decisions. Furthermore, when the final decision is wrong, it is ethically vital whether the AI advisor convinced the human decision-maker to accept and follow the wrong advice or whether the human subject was not able to make a correct decision even without the presence of AI advice.

To minimize the effect of this drawback, adding the initial human decisions helps. In their study, \cite{Schemmer2023AReliance} proposed a quantifiable two-dimensional measurement concept, Appropriateness of Reliance (AoR) 
(Eq. \ref{equation:AoR}). In this equation, RSR is the relative self-reliance (Eq. \ref{equation:RSR}) and is measured as the proportion of correct self-reliance (CSR) (where the human initial decision is correct and the incorrect AI advice is ignored; hence, the final decision is correct, CSR is 1) and IAs (IA is 1 if the initial human decision is correct and AI advice is incorrect, and 0 otherwise). Similarly, RAIR is relative to AI reliance (Eq. \ref{equation:RAIR}) and is measured as the proportion of correct AI reliance (CAIR) and CAs (CA is 1 if the initial human decision is wrong and AI advice is correct and 0 otherwise). In their definition of AR, \cite{Schemmer2023AReliance} focused on the performance and defined AR as any tuple of RSR and RAIR that leads to complementary team performance (CTP).

\begin{equation}
    \label{equation:RSR}
    \text{Relative self-reliance (RSR)} = \frac{\sum_{i=0}^N \text{CSR}_i}{\sum_{i=0}^N \text{IA}_i}
\end{equation}

\begin{equation}
    \label{equation:RAIR}
    \text{Relative AI reliance (RAIR)} = \frac{ \sum_{i=0}^N CAIR_i }{ \sum_{i=0}^N CA_i }    
\end{equation}

\begin{equation}
    \label{equation:AoR}
   \text{Appropriateness of Reliance (AoR)} = (RSR ; RAIR)
\end{equation}

The results of their study validate previous research indicating that explanations affect relative AI reliance (RAIR), while no significant effect was observed on relative self-reliance (RSR). Furthermore, explanations do not reduce over-reliance in all cases, revealing a need for new techniques to enable humans to distinguish incorrect AI advice. Moreover, in the relationship between CTP and AoR, they argued that achieving CTP depends on the relationship between RAIR and RSR and the presence of complementary potential, i.e., in some scenarios of a task, AI outperforms humans. In contrast, in other cases, humans should outperform AI \cite{Schemmer2023AReliance}.




\section{XAI and Decision Making}
\label{sec:xai_and_dm}

The field of AI-assisted decision-making research is expanding at a rapid pace, with a steady increase in the number of studies and experiments being conducted. However, there is a mixed set of findings on the impact of XAI on human decision-making due to various tasks and setups among experiments \cite{schemmer2022meta}. There is also a need for empirical studies to form a foundational understanding of the interaction between humans and AI in order to make decisions \cite{lai2023towards}. Furthermore, the role of humans in the explainability of the models and its impact on AI-assisted decision-making with and without XAI components is not adequately investigated \cite{adadi2018peeking}. For example, there exists a large body of research in cognitive science, social sciences, and psychology regarding how humans form explanations and explain different phenomena to each other \cite{miller2019explanation}, which can be beneficial for the field of XAI, especially in an AI-assisted decision-making setting. In this section, a review of each of the papers mentioned in table \ref{tab:decisionMakingEmpiricalStudies} is provided.

\subsection{Review}

Alufaisan et al. \cite{alufaisan2021does} found through their two-choice classification experiment on real datasets that any AI prediction provided to the participants can improve their decision-making accuracy. At the same time, there is no conclusive evidence that explanations have a meaningful impact \cite{alufaisan2021does}. Their findings demonstrate that the effect of AI condition (having AI prediction) depends on the dataset used. Moreover, the explanations offer improved participant decision-making accuracy compared to the control condition, but no significant improvement is observed compared to the AI (with no explanation) condition. Their results also indicated a positive relationship between AI mean accuracy and the participant's mean accuracy, with a small number of participants outperforming the AI accuracy\cite{alufaisan2021does}.

Previous work evaluated human-AI collaboration in decision-making situations through different factors, including accuracy, speed, and independence (reliance). While previous studies observed complementary improvements in the presence of explanations, the scenarios were limited to AI outperforming human and team accuracy. Bansal et al. \cite{bansal2021does} conducted an empirical study to gauge the impact of explanations on achieving higher accuracy for the human-AI team compared to the individual performance of the human and AI, where human and AI performance are comparable. They used two tasks, a text classification (sentiment analysis) and a question-answering task, on three datasets to amplify the generalizability of the results. In terms of explanations, they incorporated three strategies: (1) explaining the predicted class, (2) explaining the top two predicted classes, and (3) a dynamic explanation strategy that chooses between the other two strategies based on AI confidence. The dynamic approach chooses strategy (2) only when the AI confidence is below a model and task-specific threshold. While their findings showed that providing the model confidence achieved complimentary performance, there was no evidence that explanations improved the team performance.
Furthermore, in cases where the AI is incorrect, providing explanations worsens the decision accuracy. Finally, the proposed adaptive explanation strategy successfully reduced user reliance on AI as participants showed less tendency to agree with the model recommendation where the model confidence did not pass the adaptive threshold.

In another study of text classification of online toxicity, Carton et al. \cite{carton2020feature} conducted a user study with six conditions among 480 subjects. The study had two phases: In phase one, participants were asked to label social media comments based on the level of toxicity. In the second phase, subjects were to predict the consensus of other annotators on a set of comments. The authors find that feature-based explanations, i.e., interpretability techniques that explain which features of input had what impact on the model output from that input \cite{murdoch2019interpretable}, do not improve accuracy or agreement with model predictions. However, explanations change the distribution of subject error so that providing explanations increases the false negative rates and decreases the false-positive rates compared to the control (un-assisted) condition. Moreover, the results showed that while providing model predictions increased the decision time for humans due to the fact that subjects needed to digest more information, adding the explanations reduced this added time penalty, bringing down the average time for comment labeling back to the control condition. Finally, the authors highlight the expectation gap between the excitement around the field of XAI and what the field has demonstrated in terms of human-AI performance.

A novel paradigm of human-AI collaboration, namely "conditional delegation," is proposed in the work of Lai et al. \cite{lai2022human}. The authors discussed that full automation of AI models in high-stakes situations, in addition to safety, ethical, and legal reasons, is undesirable due to the problem of distribution shift. Distribution shift refers to the drop in the model's performance due to facing out-of-distribution examples, i.e., the examples different from the training set. Given the situation, using conditional delegation, humans and AI work together to identify the trustworthy regions of the model before deployment. After deployment, based on whether the input is within the model's trustworthy region or not, the model can make a decision (rely on the model's outcome), or other actions can be taken, such as a manual rule-based approach or using other models that cover the untrusted regions of the first model. In their work, the authors use content moderation as a testbed to test their proposed paradigm. The results from their experiment show that conditional delegation outperforms the model when working alone.
On the other hand, achieving higher team performance depends on the distribution (in or out-of-distribution), where out-of-distribution AI shows a significant decrease in performance, which conditional delegation may not be sufficient to overcome. In terms of explanations, their findings confirmed previously mentioned works, showing no significant effect of explanations on the performance of conditional delegation. Moreover, due to the findings, global explanations negatively affect human-AI performance in union precision. Finally, for future direction, the authors recommend utilizing techniques that exploit the impact of explanations on precision through the incorporation of uncertainty and confidence and mitigating bias from the priming role of explanations.

The work of Zhang et al. \cite{zhang2020effect} emphasizes the importance of trust calibration in human-AI collaboration in high-stakes scenarios, where the expertise of humans if applied appropriately, can lead to optimized team outcomes. The success of these scenarios, however, relies on the calibration of human trust in the AI. In their study, they conducted two experiments to investigate the impact of showing confidence scores on trust calibration and AI-assisted prediction accuracy and the effect of local explanations, respectively. To measure trust, rather than measuring self-reported trust, which may not be reliable, they used two behavioral indicators: One: the percentage of using AI prediction among the situations that there is a disagreement between the human and AI (or the percentage of entrusting AI to make the decision among all trials for the condition where AI prediction is absent). Two: the percentage of situations where the human's final decision disagrees with the AI. The authors also investigate the effect of having a full versus partial model. The partial model provides less information (taking out the most influential feature from the training set), giving the human leverage of access to more information. They used forced choice decision-making to decide, given a set of demographic and job-related features, whether an individual's income exceeds a certain amount. The results indicate that when the confidence score is present, participants switch more to AI's decision. Intuitively, when the confidence level of the model is low, the switch percentage decreases compared to higher model confidence scores. Although confidence scores improved trust calibration, they had no significant impact on AI-assisted accuracy. Finally, showing explanations does not exhibit better trust calibration compared to the baseline. Their work highlights the limitation of using confidence scores and suggests the investigation of more explanation techniques for trust calibration in future work. 

Schemmer et al. \cite{Schemmer2023AReliance} conducted a study to evaluate the impact of explanations on their proposed measure to evaluate appropriate reliance (AR), namely Appropriateness of Reliance (AoR). The authors use participants' initial decisions with a judge-advisor-system concept as the study design. Using a deceptive hotel review classification for the task, the participants make an initial decision, then are provided with AI advice and are offered to update their decision. They incentivize the study, finding that explanations reduce under-reliance, highlighting the positive impact of feature importance on human-AI decision-making despite not meeting CTP. In terms of trust, they find no effect of the explanations on trust, although the findings confirm the positive correlation between trust and relative AI reliance (RAIR). The authors underscore the investigation of other types of explanations, such as global and counterfactual explanations, for future work. AR and AoR are discussed in section \ref{sec:decisionMaking}

In a similar effort to investigate the role of explanations on reliance, Vasconcelos et al. \cite{vasconcelos2023explanations} conducted five studies on a maze task using a proposed cost-benefit framework to analyze engaging with the AI explanation where the costs and benefits of engaging with the task are weighed against the costs and benefits of verifying the AI's prediction. The participants are asked to find the correct exit from the four provided exits in various mazes. Mazes vary in terms of difficulty by their size, the easiest being a grid of 10x10 to the hardest, which is a grid of 50x50. The authors argue that over-reliance is not merely the result of miscalibrated trust or cognitive biases but is rather a strategic choice by humans. They validate this claim by calculating subjective utility, which reveals that people attach higher utility to more complex tasks compared to more manageable tasks. The results of the study validate their proposed framework. According to their findings, costs such as task difficulty, explanation difficulty, and benefits such as monetary compensation can affect over-reliance. Their studies have shown that explanations have the potential to reduce over-reliance on AI in complex problems. They have also revealed that the type of explanation being used is essential, as explanations that make AI mistakes more obvious tend to reduce overreliance to a greater extent.  Increasing the benefit also improves the reduction of over-reliance

Leitchmann et al. \cite{leichtmann2023effects} use a novel test case, edible vs. poisonous mushroom classification, to investigate the role of explanations on human-AI decision-making in high-risk tasks. In a 2x2 online study, participants decide on a set of mushroom photos whether the mushrooms are edible or not. The authors use two types of explanations: feature-based and example-based. They also investigate the impact of an educational intervention, i.e., providing high-level information about how ML models, including the one used in the study, work. Their findings demonstrate no effect of educational intervention and domain-specific knowledge, which the authors suspect may be due to the short amount of information provided. In contrast, the presence of explanations enhances task accuracy. In terms of trust, participants with visual explanations show significantly less trust in the model predictions, especially when the model is wrong. The authors mention the limitations of their study and emphasize the need for explainable AI methods and further research on user comprehension and trust in AI-based classification systems.

Table \ref{tab:review_summary} provides a comprehensive list of details for the empirical studies conducted in each paper for easier comparison. The following sections provide a discussion of the findings and lay out future research directions.

\begin{table}[!ht]
    \tiny
    \caption{Summarized details of the surveyed papers}
     \label{tab:review_summary}
    \begin{tabularx}{\linewidth}{X X X X X X X X}
      
    \toprule 						
        \textbf{Reference} & \textbf{Independent Variables} & \textbf{Dependent Variables} & \textbf{Task} & \textbf{Data} & \textbf{XAI method} & \textbf{AI model} & \textbf{Subject Pool}  \\
        \midrule
         \addlinespace
         Leichtmann et al\.\, 2023 \cite{leichtmann2023effects} & Explanation (attribution-based, example based) Educational Intervention & Human classification performance, trust, AI comprehension & classification: edible/ poisonous mushrooms, take home/not take home mushrooms &	\cite{deng2009dataset} & attribution-based (GradCAM) Example-based (ExMatchina*) & ResNet50* \cite{he2016resnet} & 410 \\
         \addlinespace
         Schemmer et al\.\, 2023 \cite{Schemmer2023AReliance} & Feature-based explanations & Appropriateness of Reliance (AoR) , Confidence & deceptive hotel review classification	& \cite{ott2011finding, 
         ott2013negative} & LIME \cite{ribeiro2016should}: Text highlighting & SVM & 299\\
         \addlinespace
         Vasconcelos et al\.\, 2023 \cite{vasconcelos2023explanations} & Cost manipulation: Task difficulty (study1) , Explanation difficulty (Study 2,3), Benefit manipulation: Monetary bonus (Study 4), Utility of XAI methods (Study 5) & Over-reliance, Need for Cognition Score &Maze solving & code-generated mazes & "Highlighting: Visual, Written: textual explanation & \- & 731 \\       
        \addlinespace
         Lai et al\.\, 2023 \cite{lai2022human} & Explanations: local explanations, global explanations. & Efficacy, efficiency and engagement & content moderation & "Wikipedia comments \cite{wulczyn2017exMachina}
         Hate speech on Reddit \cite{qian2019benchmark}" & Local explanations: Highlighting text, Global explanations & rationale-style neural architecture \cite{lei2016rationalizing} & 240  \\
        \addlinespace
         Bansal et al\.\, 2021 \cite{bansal2021does} & Explanations & Team performance: accuracy of final decisions & Task1 and 2: Sentiment analysis of book and beer reviews
         Task 3: Question answering: LSAT questions (logical reasoning) & Beer reviews \cite{mcauley2012learning}
         Book reviews \cite{he2016ups}, LSAT model fine-tuning: \cite{yu2020reclor} & Local saliency explanations: Inline highlights (Task 1 and 2): LIME  \cite{ribeiro2016should}, Narrative explanations (Task 3) & RoBERTa-based classifier \cite{liu2019roberta} & over 1500 (100 per condition)   \\ 
        \addlinespace
         Alufaisan et al\.\, 2021 \cite{alufaisan2021does} & Explanations & Objective human decision accuracy	Predict individual income, Recidivism & COMPAS \cite{dominguez2019data}, Census Income \cite{dheeru2017uci} & anchor-LIME \cite{ribeiro2018anchors} &	COMPAS: SVM with rbf kernel & Census Income: Multi layer perceptron neural network & 300 \\
        \addlinespace
         Carton et al\.\, 2020 \cite{carton2020feature} & feature attribution-style explanations,Model prediction & Accuracy, Agreement, False negative rate, False positive rate, Speed (second /comment) & Toxicity of social media posts & Wikipedia comments \cite{wulczyn2017exMachina} & Inine highlighting & LSTM & 480 \\
        \addlinespace
        Zhang et al\.\, 2020 \cite{zhang2020effect} & Experiment 1: Confidence score, Experiment 2: Local explanations & AI-assisted accuracy, Switch rate, Agreement rate & Income prediction & Adult Data Set  \cite{dheeru2017uci} & local explanations: Shapley \cite{lundberg2017unified} & Gradient Boosting Decision Tree	Experiment 1: 72, Experiment 2: 9 \\
        \addlinespace
        \bottomrule
    \end{tabularx}

\end{table}

\section{Discussion}
\label{sec:discussion}
In addition to the limitations of the current work, this section provides a summary of challenges of the application of explainable artificial intelligence to AI-assisted decision making found during the literature review and provides future direction for researchers.

\subsection{Challenges}
The survey of several studies highlights the challenges in conducting human-subject experiments to investigate the impact of XAI methods on human-AI decision-making. This section covers some of these challenges:

The review of the literature on empirical human-AI decision-making reveals mixed outcomes. While some studies demonstrate the effectiveness of explanations in increasing team performance (and in some cases achieving complimentary team performance) \cite{lai2019human, lai2020chicago, zhang2020effect} or reducing over-reliance \cite{zhang2020effect}, others show no effect of explanations in comparison to AI-only conditions \cite{bansal2021does, lai2022human, poursabzi2021manipulating}. The surveyed works in this paper suggested factors that may clarify these discrepancies in results:

\textbf{Human performance profile}
Accounting for human performance profiles in designing human-AI decision-making experiments can be vital. The application of this consideration can be in developing the model where the performance of the model should not be far from human performance. The concept may also be applied to task design, where the error boundaries of humans and AI are complementary and not aligned, meaning that where AI tends to err, humans should not make a mistake. Controlling such performance profiles may help achieve team performances beyond the performance of AI or humans individually \cite{zhang2020effect}. 

\textbf{Task difficulty}
The level of task difficulty can affect the effectiveness of explanations. If the task is too easy for the human, the task could be accomplished alone \cite{zhang2020effect}. On the other hand, \cite{alufaisan2021does} suggests that rather than task difficulty level, how intuitive the dataset for the task is may be the contributing factor. It opens a venue for future research.

\textbf{Experiment design}
The design of the experimental setup for empirical studies is crucial. For instance, the order in which explanations are presented to human participants can have a significant impact on the results, as evidenced in \cite{Schemmer2023AReliance}. Studies have shown that conducting the task first and then revealing AI explanations can mentally prepare participants and potentially change their behavior, as described in \cite{buccinca2021trust}. This phenomenon could explain the lower relative reliance on AI observed in a study whose reference is unknown. Sequential task setup can also lead to an anchoring effect, which reduces human engagement with AI, as reported in \cite{buccinca2021trust}.
Future studies can investigate different setups, namely sequential versus non-sequential, to measure the change in engagement with AI (RAIR).

\textbf{Interpretability methods}
Research shows that the application of explanation methods is not always beneficial. Bansal et al. \cite{bansal2021does} find a concerning observation that explanations contributed to blind trust rather than reducing over-reliance. The authors suggest the development of explanations based on the frequency of agreement between the human and the model. Lai et al. \cite{lai2022human} points out another challenge. The use of global explanations in their work, the generated top tokens in explanations do not necessarily correlate with high precision. The recommended techniques consider confidence as a proxy for generating explanations.
Future research can explore various techniques and methods of explanation to address such challenges.

\textbf{Visualization}
Recent research shows that data visualization techniques can affect user judgments. Karduni et al. \cite{karduni2021bayesian} showed that communication of uncertainty using their proposed method of "Line+Cone" can affect users' belief update and confidence in correlation judgment of bi-variate data.
In another study, Markant et al. demonstrate that data visualizations can update users' beliefs about a specific topic. However, the amount of change may be affected by the users' broader attitude about the issue \cite{markant2023whendo}. The effect of applying different visualization techniques, such as interactivity and uncertainty communication methods, on team performance in XAI methods can be further investigated.

\subsection{Limitations}
With a focus on the empirical study of the role of explainable artificial intelligence (XAI) in human decision-making, this paper surveyed a non-exhaustive list of papers. The field of XAI has received much attention in recent years with the emergence of complex deep-learning model architectures. With an abundance of works contributed to this area, this paper only considers a handful of papers from 2019 onward, which may leave a number of highly praised works out.
Another limitation of this work is the mere focus on an empirical study of the role of XAI in AI-assisted decision-making. Although empirical research with human subjects is essential for 'human in the loop' systems, a survey of theoretical works could benefit the understanding of the challenges and opportunities in achieving complimentary team performance in human-AI decision-making.

\section{Conclusion and Future Direction}
\label{sec:future directions}
High-performance deep neural networks are taking various domains by storm. The high performance of these models, however, comes at a price: more complexity and less transparency of the models, which makes them less understandable and, hence, less trustworthy for the end users, mainly when these models are applied to high-stakes domains. In addition, the AI models are only as good as the data they are trained on. More importantly, the real-world scenarios comprise unique (out-of-distribution) situations that reduce model performance drastically. All these reasons, in addition to legal, ethical, and safety concerns, provide evidence that full autonomy in high-stakes tasks is not desired. Human-AI decision-making is an interdisciplinary area of research, multifaceted with concepts from Human-Computer Interaction (HCI), Machine Learning, Cognitive Sciences, and data visualizations. An empirical study of human-AI decision-making constitutes components including task design, model development, and choice of explanation methods, each requiring a thorough investigation to achieve a complementary team performance.
This paper paints a picture of the application areas of human-AI decision-making systems, as well as provides a brief overview of explainable artificial intelligence methods. The paper discusses decision-making, including examples of decision-making models from psychology and cognitive science perspectives. Finally, a survey of selected papers conducted by an empirical study on the topic is provided, and the findings are discussed. During the review of the literature, a mixture of results on the effectiveness of explanations on enhancing team performance and reduction of over-reliance is exposed.
Future research in human-subject experiments on AI-assisted decision-making should investigate various aspects. Three suggested directions are: 1) Explanation methods: The ensamble of different explanation methods, different use cases, and the relation between explanation methods, AI model architectures, and the task. 2) Various levels of task difficulty should be tested to identify the areas where complimentary performance for the team is more likely, including consideration of human performance profile and AI and human performance relatively. 3) Investigation of visualization techniques for explanations and their impact on the team performance is another area for future study.

\bibliographystyle{unsrt}
\bibliography{references}

\clearpage

\end{document}